\documentclass[prd,letterpaper,12pt,showpacs,preprint]{revtex4}

\usepackage{graphicx}
\usepackage{amsmath}
\usepackage{bm}

\begin{document}
\preprint{ \vbox{\hbox{ JLAB-THY-05-386}}}

\title{Scaling of Dirac Fermions and the WKB approximation}

\author{J.~W.~Van Orden$^{1,2}$, Sabine Jeschonnek$^{3}$, and John Tjon$^{4,5}$ }

\affiliation{\small \sl (1) Department of Physics, Old Dominion University, Norfolk, VA 23529\\
(2) Jefferson Lab, 12000 Jefferson Ave, Newport News, VA 23606\\
(3) The Ohio State University, Physics
Department, Lima, OH 45804\\
(4) Department of Physics, National Taiwan University, Taipei
10617, Taiwan\\
(5) KVI, University of Groningen, The Netherlands }

\date{\today}

\begin{abstract}
We discuss a new method for obtaining the WKB approximation to the
Dirac equation with a scalar potential and a time-like vector
potential. We use the WKB solutions to investigate the scaling
behavior of a confining model for quark-hadron duality. In this
model, a light quark is bound to a heavy di-quark by a linear
scalar potential. Absorption of virtual photons promotes the quark
to bound states. The analog of the parton model for this case is
for a virtual photon to eject the bound, ground-state quark
directly into free continuum states. We compare the scaling limits
of the response functions for these two transitions.
\end{abstract}


\maketitle

\section{Introduction}

Quark-hadron duality implies that in certain kinematic regimes,
properly averaged hadronic observables can be described by a
perturbative QCD (pQCD) calculation. This version of duality is
highly relevant as perturbative QCD calculations can be performed.
Using duality, these pQCD calculations can then be related to
averaged data taken in the resonance region. Duality was first
observed more than 30 years ago experimentally by Bloom and Gilman
in inclusive inelastic electron scattering \cite{bgduality}, and has
since then been observed in a variety of reactions: $e^+ \, e^- \to
hadrons$ is an example for duality we know from the textbooks, the
semileptonic decay of heavy mesons is another example
\cite{richural,semilep,isgurwise}, duality is considered in the
analysis of heavy ion reactions \cite{ralf}, and forms the basis for
using QCD sum rules \cite{qcdsr}. Recently, duality has been
observed to high precision and down to rather low momentum transfers
in electron scattering at Jefferson Lab
\cite{jlab,f1fldual,prdmom,emcdual}. Duality in spin observables is
currently probed by several experiments, both at Jefferson Lab and
at DESY in Germany \cite{meziani,hermes,hermesg1,e01012}. In
addition to the ``classical'' examples and applications  of duality,
duality ideas are applied in new areas, too. For neutrino
scattering, the beam energies are not well known, and an averaging
will thus take place almost automatically. The application of
duality is discussed for several planned neutrino experiments, see
e.g. \cite{minerva}, and duality ideas have been applied in
\cite{Horowitz:2004yf} to nucleon/nuclear duality in neutrino
scattering. There is also interest in duality in parity violation
experiments \cite{vipuli}, and with regard to generalized parton
distributions \cite{closezhaogpd,diehl}. A very local version of
duality - assuming that it holds for just one resonance -  has been
used in \cite{wallymprl,adelaide} to extract information on
structure functions at $x_{Bj} \to 1$ in the scaling limit from form
factor data. These ideas were also applied to neutrino-nucleon
scattering \cite{adelaide}. Duality ideas might also be useful for
pion photoproduction \cite{haiyan}. Duality is a major point in the
12 GeV upgrade of CEBAF at Jefferson Lab \cite{12gevwp}. A recent
review of quark-hadron duality can be found in \cite{mek}.

Apart from being interesting all by itself, quark-hadron duality is
an important tool for studying kinematic areas that cannot be
accessed in the deep inelastic regime: measurements at large values
of $x_{Bj}$ in the resonance region typically have much higher count
rates than measurements at the same $x_{Bj}$ in the deep inelastic
regime, which requires very high four-momentum transfers $Q^2$.
Application of a proper averaging procedure to the resonance data
may allow the extraction of deep inelastic information, e.g. in the
case of the polarization asymmetry $A_1$ of the neutron for $x_{Bj}
\to 1$. Before duality averaging procedures can be applied safely,
we require a thorough understanding of where duality holds and how
exact it is. Therefore, duality has been studied intensively by
theorists during the past couple of years
\cite{closeisgur,closewallym,closezhao,closezhaogpd,carlnew,carlnimay,
donghe,dongli,morechinese, pp,mpdirac,marka1,kiev, leyaouancped,
simula,wallymprl, adelaide,ijmvo,jvod2,dirac,poldirac,elba,myhugs,
zoltanfranz,liuti,davidovsky,hofmann,lee}. Most of the theoretical
studies focus on duality in electron scattering, due to the large
experimental program. The model we will discuss here is for electron
scattering, too.

Many recent papers have been devoted to modeling quark-hadron
duality in simple, fully solvable relativistic models, to gain a
better understanding of the conditions under which duality works
\cite{closezhao,closezhaogpd,pp,mpdirac,marka1,ijmvo,jvod2,dirac,poldirac}.
The idea of modeling duality is to capture just the essential
physical conditions of this rather complex phenomenon. Typically,
these basic requirements for a model are imposed: one requires a
relativistic description of confined valence quarks, and one
treats the hadrons in the infinitely narrow resonance
approximation.

In these papers, one important point was raised and discussed that
is interesting not just for duality, but in general: do the
scaling curves obtained assuming outgoing plane waves agree with
the scaling curves obtained when we assume final state
interactions?

The general approach is to model the perturbative QCD (pQCD) picture
by considering a quark bound within a potential in the initial
state, and after the interaction with a virtual photon, the quark is
considered ``free'' and the potential set to zero. This
``bound-free'' transition corresponds to a plane wave impulse
approximation (PWIA). This transition is compared with a hadronic
picture: the initial state is the same - the quark is bound in a
potential - but after the absorption of the photon, it is in an
excited, but still bound, state. This ``bound-bound'' transition
corresponds to the excitation of resonances.

In order to reproduce duality, a model must fulfill several
conditions that are observed experimentally: first, the bound-free
transition, corresponding to pQCD, must scale for large momentum
transfers. Second, the bound-bound transition must scale, and the
bound-bound and bound-free scaling functions have to agree so that
the third condition can be fulfilled: at low momentum transfers, the
bound-bound results should oscillate around the bound-free scaling
curve. Note that in all models currently proposed, there are no
gluons included, and therefore neither radiative corrections nor
evolution of scaling curves are present.

In our recent papers \cite{ijmvo, jvod2}, we could show analytically
that the two different scaling curves, found for the bound-free and
bound-bound transition, do coincide. In these papers, we considered
a model where all particles were treated as scalars \cite{ijmvo},
and a model where only the quarks were treated as scalars, while
electrons and photons had their proper spin \cite{jvod2}. The
treatment of the quarks as scalars considerably simplified our
calculations, as the resulting Klein-Gordon equation could be solved
analytically, by recognizing that it could be rewritten to resemble
a Schr\"odinger equation. Once we introduced the proper spin for
quarks \cite{dirac,poldirac}, we solved the Dirac equation
numerically, using a Runge-Kutta-Fehlberg (RKF) algorithm
\cite{Fehlberg}. The numerical accuracy we achieved this way
restricted us to momentum transfers below $q < 12 GeV$, a value
where the calculated response functions had not fully converged to
their scaling value yet. Even though we solved only for
eigenenergies of up to $12 GeV$, we found roughly 24,000 states
below that energy. As for any relativistic problem, the density of
energy states rapidly increases with the energy.

Thus, we could not determine if the bound-bound and bound-free
scaling curves were going to coincide, as for the other,
simplified models we investigated previously.

In a recent series of papers by Paris et al.
\cite{pp,mpdirac,marka1}, similar models were considered, and
solved numerically, and the result found there was a pronounced
discrepancy between the bound-free and bound-bound transition.
This led the authors to question the interpretation of the scaling
curves extracted from deep inelastic scattering (DIS). In DIS, it
is generally assumed that final state interactions are negligible,
and that the scaling curves can be interpreted in terms of parton
distribution functions.

In this paper, we investigate the scaling behavior of our model with
Dirac quarks. We do this by employing the WKB method to solve the
Dirac equation numerically for very high momentum transfers. This
allows us to investigate the scaling behavior of the response
functions in the bound-bound transition at the relevant high
momentum transfers. The bound-free transition does not require any
complicated final state calculations, and can be evaluated in a
straightforward manner.

This paper is organized as follows: first, we remind the reader of
the general ideas underlying the WKB approximation\cite{WKB}, and
discuss the WKB for solving the Dirac equation. Then, we introduce
our model and present numerical results obtained in the WKB
approximation. We compare them to the results obtained by
explicitly solving the differential equation at $q = 10~GeV$, the
highest value accessible with the Runge-Kutta-Fehlberg method.
After validating the WKB calculations, we proceed to investigate
the scaling behavior of the bound-bound and bound-free transitions
at large $q$, where convergence has set in.

\section{The WKB Approximation for the Dirac Equation}

The Dirac hamiltonian for a particle in a scalar filed $S(r)$ and a
time-like vector field $V(r)$ is
\begin{equation}
\hat{H}=\bm{\alpha}\cdot\hat{\bm{p}}+\beta\left(m+S(r)\right)+V(r)
\,.
\end{equation}
The wave functions for the Dirac hamiltonian are
\begin{equation}
\psi_{nljm}(\bm{r})=\left(
\begin{array}{c}
\frac{G_{nlj}(r)}{r}{\cal Y}^m_{lj}(\Omega_r)\\
i\frac{F_{nlj}(r)}{r}{\cal Y}^m_{\overline lj}(\Omega_r)
\end{array}
\right)
\end{equation}
where the ${\cal Y}^m_{lj}(\Omega_r)$ are the usual spin spherical
harmonics. Then angular momentum quantum numbers are $l=-\kappa+1$
for $\kappa<0$ and $l=\kappa$ for $\kappa>0$ with $\overline
l=-\kappa$ for $\kappa<0$ and $\overline l=\kappa-1$ for
$\kappa>0$. The reduced radial wave functions $G(r)$ and $F(r)$
are solutions to the coupled equations
\begin{eqnarray}
\hbar G'(r)+\hbar\frac{\kappa}{r}G(r)&=&(m+S(r)-V(r)+E)F(r)\label{diff1}\\
\hbar F'(r)-\hbar\frac{\kappa}{r}F(r)&=&(m+S(r)+V(r)-E)G(r)
\,.\label{diff2}
\end{eqnarray}

The differential equation can be solved numerically, but above a
certain energy, finding the numerical solution with a standard
Runge-Kutta-Fehlberg method \cite{Fehlberg,numrecipes} becomes
inaccurate. In the relativistic treatment of a potential, the energy
eigenvalues of higher states lie closer and closer together. At some
high enough energy, the separation between neighboring states
becomes so small that some states may be missed with shooting
methods. Also, one always has to calculate all states consecutively,
instead of being able to calculate a state with certain, given
quantum numbers. While an improved method for the integration of the
differential equation may be applied, the situation lends itself to
the application of a semiclassical approximation, the WKB method
\cite{WKB}.

\begin{figure}
\centerline{\includegraphics[height=4.0in]{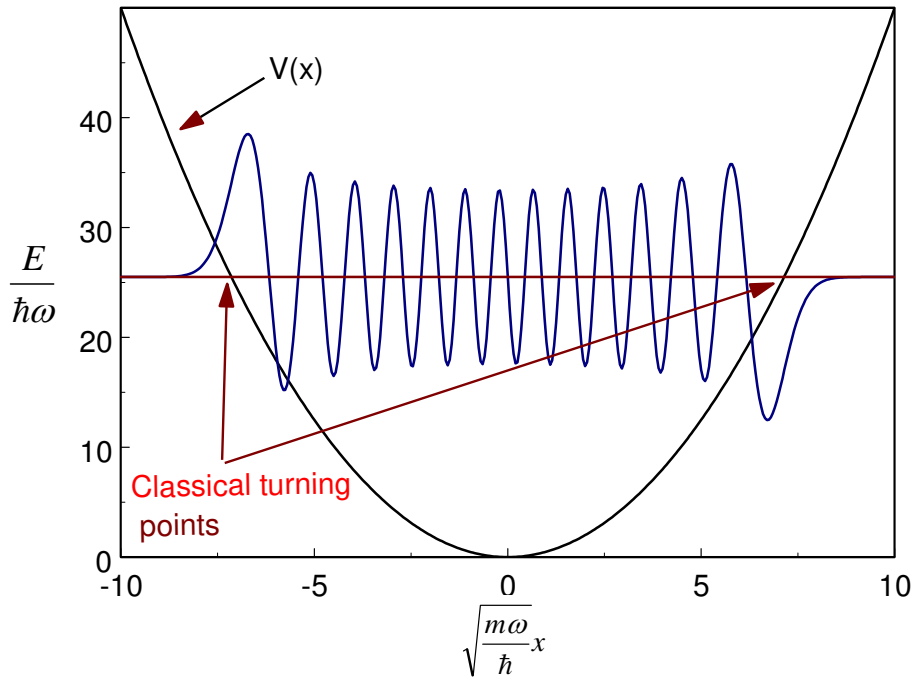}}
\caption{The potential of a harmonic oscillator is shown, together
with the exact wave function for an energy of $E = 26 \hbar
\omega$.} \label{figharm}
\end{figure}
In order to illustrate that the assumptions of the WKB
approximation are perfectly reasonable, we show a plot of a
one-dimensional harmonic oscillator potential in
Fig.~\ref{figharm}, together with the exact solution of the wave
function for an energy $E = 26 \hbar \omega$. One can see that the
wave function oscillates in the classically allowed region, where
$E > V$, and strongly resembles a plane wave there. Outside of the
classically allowed region, the wave function is damped and goes
to zero.

The WKB approximation can be applied when the potential has only a
small variation over several wavelengths of the particle, as is
the case in Fig. \ref{figharm} near the origin. This feature makes
it quite useful for studying the highly energetic excited states
of a particle bound in a potential. The WKB approximation assumes
that in the classically allowed regions, the wave function can be
approximated by a plane wave with a position dependent effective
wave vector. As the potential is almost constant over a few
wavelengths, the change of the wave vector is small compared with
the value of the local wave vector.

The WKB approximation can then be viewed as an expansion in
derivatives of the effective phase. In the case of a one-dimensional
wave equation, this is equivalent to an expansion in $\hbar$ and is,
therefore, often referred to as the semiclassical or quasiclassical
approximation.  In the case of the radial equation for a
three-dimensional Schr\"odinger wave equation, care must be taken to
correctly treat the centrifugal barrier term that arises from the
angular momentum operator acting on the angular functions. This term
is also important for the solution of the corresponding classical
problem, and needs to be included in the leading-order effective
potential for the WKB approximation, even though it will involve
powers of $\hbar$. The WKB approximation is then carried out by
attaching an expansion parameter to the derivative terms in the wave
equation and then proceeding as in the one-dimensional case.

%{\bf this needs work to make it more clear}

To implement the WKB approximation for the Dirac equation the
radial equations (\ref{diff1}) and (\ref{diff2}) are modified by
making the substitution $\hbar\kappa\rightarrow\kappa$ and
replacing the remaining factors of $\hbar$ by the expansion
parameter $\eta$ to give
\begin{eqnarray}
\eta G'(r)+\frac{\kappa}{r}G(r)&=&(m+S(r)-V(r)+E)F(r)\label{diff1a}\\
\eta F'(r)-\frac{\kappa}{r}F(r)&=&(m+S(r)+V(r)-E)G(r)
\,.\label{diff2a}
\end{eqnarray}
We now can parameterize the wave functions for positive energy
solutions as
\begin{equation}
G(r)=e^{\frac{i}{\eta}\xi(r)}\label{GWKB}
\end{equation}
and
\begin{equation}
F(r)=B(r)e^{\frac{i}{\eta}\xi(r)}\label{FWKB}
\end{equation}
where $\xi(r)$ is the local phase for a plane wave and $B(r)$
allows for differences in the upper and lower component wave
functions. Substituting these into (\ref{diff1a}) and
(\ref{diff2a}) yields
\begin{eqnarray}
i\xi'(r)+\frac{\kappa}{r}&=&(m+S(r)-V(r)+E)B(r)\label{diff1b}\\
\eta B'(r)+iB(r)\xi'(r)-\frac{\kappa}{r}B(r)&=&(m+S(r)+V(r)-E)
\,.\label{diff2b}
\end{eqnarray}
Equation (\ref{diff1b}) can be solved to give
\begin{equation}
B(r)=\frac{i\xi'(r)+\frac{\kappa}{r}}{m+S(r)+V(r)-E}\,.
\label{Bofr}
\end{equation}
Substituting (\ref{Bofr}) into (\ref{diff2b}) produces the second
order differential equation for the phase
\begin{eqnarray}
&&i\eta\xi''(r)-{\xi'(r)}^2-i\eta\frac{S'(r)-V'(r)}{m+S(r)-V(r)+E}\,\xi'(r)
-\frac{\kappa^2}{r^2}-(m+S(r))^2+(V(r)-E)^2\nonumber\\
&&\qquad-\eta\frac{\kappa}{r}\left(\frac{1}{r}+
\frac{S'(r)-V'(r)}{m+S(r)-V(r)+E}\right)=0\,.\label{WKB1}
\end{eqnarray}

The WKB approximation is obtained by expanding the phase function
in powers of $\eta$ as
\begin{equation}
\xi(r)=\sum_{n=0}^\infty \eta^n \xi_n(r)\label{phaseexp} \,.
\end{equation}
Substituting (\ref{phaseexp})  into (\ref{WKB1})  gives
\begin{eqnarray}
&& i \sum_{n=1}^\infty \eta^n
\xi_{n-1}''(r)-\sum_{n=0}^\infty\eta^n\sum_{m=0}^n\xi_m'(r)\xi_{n-m}'(r)
-\frac{S'(r)-V'(r)}{m+S(r)-V(r)+E}\,\sum_{n=1}^\infty\eta^n
i\xi_{n-1}'(r)\nonumber\\
&&-\frac{\kappa^2}{r^2}-(m+S(r))^2+(V(r)-E)^2-\eta\frac{\kappa}{r}\left(\frac{1}{r}+
\frac{S'(r)-V'(r)}{m+S(r)-V(r)+E}\right)=0\,.
\end{eqnarray}
Equating coefficients of like powers of $\eta$ gives
\begin{equation}
-{\xi_0'}^2(r)-\frac{\kappa^2}{r^2}-(m+S(r))^2+(V(r)-E)^2=0\label{xi0a}
\end{equation}
for $n=0$, and
\begin{equation}
i\xi_0''(r)-2\xi_0'(r)\xi_1'(r)-\frac{S'(r)-V'(r)}{m+S(r)-V(r)+E}i\xi_0'(r)
-\frac{\kappa}{r}\left(\frac{1}{r}+
\frac{S'(r)-V'(r)}{m+S(r)-V(r)+E}\right)=0\label{xi1a}
\end{equation}
for $n=1$.
% and
%\begin{eqnarray}
%&&\ i \xi_{n-1}''(r)-\sum_{m=0}^n\xi_m'(r)\xi_{n-m}'(r)
%-\frac{S'(r)-V'(r)}{m+S(r)-V(r)+E}\, i\xi_{n-1}'(r)=0\label{xina}
%\end{eqnarray}
%for $n>1$.

Equation (\ref{xi0a})  can be solved to give
\begin{equation}
 \xi'_0(r)=\pm\sqrt{(V(r)-E)^2-(m+S(r))^2-\frac{\kappa^2}{r^2}}\equiv\pm
k_0(r)\label{xi0b}
\end{equation}
and (\ref{xi1a}) can be solved to give
\begin{eqnarray}
\xi'_1(r)&=&\frac{i}{2}\,\frac{\xi_0''(r)}{\xi_0'(r)}-\frac{i}{2}\,
\frac{S'(r)-V'(r)}{m+S(r)-V(r)+E}-\frac{\kappa}{2\xi_0'(r)r}\left(\frac{1}{r}+
\frac{S'(r)-V'(r)}{m+S(r)-V(r)+E}\right)\nonumber\\
&=&i\Im(\xi_0'(r))\pm k_1(r)\,,\label{xi1b}
\end{eqnarray}
where
\begin{equation}
k_1(r)\equiv-\frac{1}{2k_0(r)}\frac{\kappa}{r}\left[\frac{1}{r}+\frac{S'(r)
-V'(r)}{m+S(r)-V(r)+E}\right] \label{k1}
\end{equation}
and
\begin{eqnarray}
\Im(\xi_0'(r))&=&\frac{1}{2}\,\frac{\xi_0''(r)}{\xi_0'(r)}-\frac{1}{2}\,
\frac{S'(r)-V'(r)}{m+S(r)-V(r)+E}\nonumber\\
&=&\frac{d}{dr}\ln(\sqrt{k_0(r)})
-\frac{d}{dr}\ln(\sqrt{m+S(r)-V(r)+E})\,.\label{Imxi1}
\end{eqnarray}

 Keeping terms to order $\eta$, and using (\ref{xi0b}) and
(\ref{xi1b}) the phase function can be integrated to give
\begin{eqnarray}
\xi(r)&=&\int^r dr'\left(\xi_0'(r')+\eta
\xi_1'(r')\right)\nonumber\\
&=&\int^r dr'\left[\pm
k_0(r')+i\eta\frac{d}{dr'}\ln(\sqrt{k_0(r')})-i\eta\frac{d}{dr'}\ln(\sqrt{m+S(r')-V(r')+E})
\pm\eta k_1(r')\right]\nonumber\\
&=&i\eta\ln(\sqrt{k_0(r)})-i\eta\ln\left(\sqrt{m+S(r)-V(r)+E}\right)\pm\int^r
dr'\left(k_0(r')+\eta k_1(r')\right)
\end{eqnarray}
The upper-component radial wave function can then be written as
\begin{equation}
G(r)={\cal
N}\frac{\sqrt{m+S(r)-V(r)+E}}{\sqrt{k_0(r)}}\exp\left[\pm\frac{i}{\eta}\int^r
dr'(k_0(r')+\eta k_1(r'))\right]\,,\label{Gofr}
\end{equation}
where $\cal N$ is the normalization constant.

The lower-component wave function can be obtained directly by
using (\ref{Bofr}) to first order in $\eta$ yielding
\begin{eqnarray}
F(r)&=&{\cal N}\frac{1}{\sqrt{k_0(r)}\sqrt{m+S(r)-V(r)+E}}
\left[\frac{\kappa}{r}-\eta\Im(\xi_0'(r)) \pm i\left(k_0(r)+\eta
k_1(r)\right)\right]\nonumber\\
&&\qquad\times\exp\left[\frac{i}{\eta}\int^r dr'(k_0(r')+\eta
k_1(r'))\right]\,.\label{Fofr}
\end{eqnarray}
From this point we will set $\eta=1$ for convenience.

In constructing the above solution we have assumed that we are
examining the solution in the classically allowed region where the
energy $E$ is greater than the effective potential. For the
potentials which we will use, there is only one interval in $r$
which is classically allowed. It is bounded by the classical turning
points $r_\pm$ which are the solutions to
\begin{equation}
k_0(r_\pm)=0\,.
\end{equation}
Quantization of the bound states is obtained by requiring that
\begin{equation}
\int_{r_-}^{r_+}dr'(k_0(r')+k_1(r'))=\left(n-\frac{1}{2}\right)\pi\,,
\end{equation}
where $n\ge 1$, as in the Schr\"odinger case. The solutions can be
extended to the classically forbidden regions by analytic
continuation. Since the $k_0(r)$ vanishes at the classical turning
points, the local wavelength $\lambda(r)=2\pi/k_0(r)$ becomes
arbitrarily large near the classical turning points in contradiction
to the basic assumption of the WKB approximation. The expansion in
$\eta$ therefore does not converge in the vicinity of the classical
turning points. Techniques for matching the wave functions at the
turning points and replacing them with smooth approximate wave
functions near the turning points are well described in most
introductory graduate quantum mechanics texts. There is, however,
one additional complication in this case associated with the
first-order phase  $\xi_1(r)$ which does not appear in the
Schr\"odinger case. This requires us to modify the usual approach to
approximating the wave functions near the classical turning points.
The solution to this complication is discussed in Appendix A.

We will need to construct the wave functions in three regions: the
classically forbidden region where $0\leq r< r_-$ (Region I); the
classically allowed region where $r_-\leq r\leq r_+$ (Region II);
and the classically forbidden region where $r_+< r$ (Region III).
The wave functions must be matched at the boundaries. The
procedure for the upper-component wave function is the same as for
the Schr\"odinger case.
%That is the wave function in Region II is constructed in the following way.
Assume that we start by choosing the root in Eq. (\ref{xi0b}) where
$\xi_0'(r)=+k_0(r)$. The wave function is then obtained by choosing
\begin{equation}
{\cal N}=\frac{N}{2}e^{-i\frac{\pi}{4}}\label{matchPhase}
\end{equation}
and then defining the wave function as
\begin{equation}
G_{II}(r)=G(r)+G^*(r)\,.
\end{equation}
This gives
\begin{equation}
G_{II}(r)=N\frac{\sqrt{m+S(r)-V(r)+E}}{\sqrt{k_0(r)}}\cos\left(\int_{r_-}^r
dr'(k_0(r')+ k_1(r'))-\frac{\pi}{4}\right)\,.\label{GII}
\end{equation}
Applying the same procedure to the lower component yields
\begin{eqnarray}
F_{II}(r)&=&N\frac{1}{\sqrt{k_0(r)}\sqrt{m+S(r)-V(r)+E}}\nonumber\\
&&\times\left[\left(\frac{\kappa}{r}-\Im(\xi_0'(r))\right)\cos\left(\int_{r_-}^r
dr'(k_0(r')+
k_1(r'))-\frac{\pi}{4}\right)\right.\nonumber\\
&&\left. \mp \left(k_0(r)+k_1(r)\right)\sin\left(\int_{r_-}^r
dr'(k_0(r')+ k_1(r'))-\frac{\pi}{4}\right)\right]\,.\label{FII}
\end{eqnarray}

In the classically forbidden regions the wave vectors become complex with
\begin{equation}
-ik_0(r)=\tilde{k}_0(r)=\sqrt{\frac{\kappa^2}{r^2}+(m+S(r))^2-(V(r)-E)^2}
\end{equation}
and
\begin{equation}
-ik_1(r)=\tilde{k}_1(r)=\frac{1}{2\tilde{k}_0(r)}\left[\frac{\kappa}{r^2}
-\frac{\kappa}{r}\frac{S'(r)- V'(r)}{m+S(r)- V(r)+ E}\right]\,.
\end{equation}
The wave functions will be either growing exponentially or
exponentially damped.  Since we require that the wave functions be
regular at the origin and damped at $\infty$, we will construct
the wave functions in the forbidden regions so that they fall off
as $r$ moves away from the turning points. Choosing
\begin{equation}
{\cal N}=\frac{N}{2}\,,
\end{equation}

The solutions in Region I require that the sign of the root
$\tilde{\xi}_0'(r)=-\tilde{k}_0(r)$ be chosen so that the wave
function vanishes at the origin. The radial wave functions in this
region are
\begin{equation}
G_I(r)=\frac{N}{2}\frac{\sqrt{m+S(r)-V(r)+E}}{\sqrt{\tilde{k}_0(r)}}\exp\left[-\int^{r_-}_r
dr'(\tilde{k}_0(r')+ \tilde{k}_1(r'))\right]\label{GI}
\end{equation}
and
\begin{eqnarray}
F_I(r)&=&\frac{N}{2}\frac{1}{\sqrt{\tilde{k}_0(r)}\sqrt{m+S(r)-V(r)+E}}
\left(\frac{\kappa}{r}-\Im(\tilde{\xi}_1'(r))+\tilde{k}_0(r)+\tilde{k}_1(r)\right)
\nonumber\\
&&\qquad\times\exp\left[-\int^{r_-}_r dr'(\tilde{k}_0(r')+
\tilde{k}_1(r'))\right]\,.\label{FI}
\end{eqnarray}

The solutions in Region III are
\begin{equation}
G_{III}(r)=(-1)^{n-1}\frac{N}{2}\frac{\sqrt{m+S(r)-V(r)+E}}{\sqrt{\tilde{k}_0(r)}}
\exp\left[-\int^r_{r_+} dr'(\tilde{k}_0(r')+
\tilde{k}_1(r'))\right]
\end{equation}
and
\begin{eqnarray}
F_{III}(r)&=&(-1)^{n-1}\frac{N}{2}\frac{1}{\sqrt{\tilde{k}_0(r)}\sqrt{m+S(r)-V(r)+E}}
\left(\frac{\kappa}{r}-\Im(\tilde{\xi}_1'(r))-\tilde{k}_0(r)-\tilde{k}_1(r)\right)\nonumber\\
&&\times\exp\left[-\int^r_{r_+} dr'(\tilde{k}_0(r')+
\tilde{k}_1(r'))\right]\,.
\end{eqnarray}

It should be pointed out that this is very similar to the WKB
approximation to the Dirac equation described in \cite{Mur,Popov}.
In this previous work the WKB approximation is assumed to be in the
form of a two-dimensional spinor of amplitude functions multiplied
by the usual exponentiated phase function. Since the radial Dirac
equations only determine two functions, it was necessary to make a
choice for one of the amplitude functions that gives an
upper-component wave function identical to that obtained here. The
other two functions were obtained by solving two-dimensional matrix
equations by introducing a dual set of spinors for a nonhermitian
matrix and projecting to obtain scalar expressions for the phase
function and the remaining amplitude function. However, this
approach does not provide an expression for the first-order
corrections to the phase that come from $\xi_1(r)$ in the derivation
given here. These contributions are necessary for a simple, smooth
extrapolation of the wave functions over the regions where the WKB
approximation does not converge.

\section{Model Calculations}

Our model consists of a constituent quark bound to an infinitely
heavy di-quark and is represented by the Dirac hamiltonian for a
particle in a scalar field $S(r)$ and a time-like vector field
$V(r)$.
\begin{equation}
\hat{H}=\bm{\alpha}\cdot\hat{\bm{p}}+\beta\left(m+
S(r)\right)+V(r)\,.
\end{equation}
The scalar potential is a linear confining potential given by
\begin{equation}
S(r)=br,\qquad\qquad b=0.18{\rm GeV}^2\,.
\end{equation}
In our model, the vector potential is provided by a vector color
Coulomb potential. We will calculate for the case where the vector
color Coulomb potential is absent, that is $V(r)=0$, and where the
vector potential is the simple static Coulomb potential
%\begin{equation}
%V (r) = V_c(r)=-\frac{4}{3}\frac{\alpha_s}{r} \,.
%\end{equation}
\begin{equation}
V(r)=-\frac{\beta}{r}\,,
\end{equation}
with $\beta=0.4$. For convenience, the mass has been chosen to be
$m=0$. For these potentials it is easily shown that the Dirac WKB
wave functions have the correct functional dependence near $r=0$ for
all values of $\kappa$.

Note that we assume that the virtual photon only interacts with the
light quark, and not with the infinitely heavy di-quark. This still
allows us to gain qualitative insight into the issues of scaling and
duality, but makes a direct comparison of numerical results from our
model to experimental data impossible: our model is much closer to
electron scattering from a $B$ meson than a proton. We would like to
point out that therefore, the values obtained e.g. for the momentum
transfer at the onset of scaling should not be compared to
experimental values obtained in inclusive electron scattering from
the proton. However, the question if the bound-bound and bound-free
transitions lead to the same scaling curve can be discussed within
this model.

\begin{table}
\caption{Comparison of eigenenergies calculated with RKF and WKB
methods for a few selected states.}\label{energies}
\begin{tabular}{rrrrr}\hline\hline
n & $\kappa$ & $E_{\rm RKF}$ (GeV) & $E_{\rm WKB}$ (GeV) & $E_{\rm
RKF}-E_{\rm WKB}$\\\tableline
1  & -1  & 0.687666 &   0.6895116  & -0.001846 \\
60 & 5  & 6.697930  &  6.6979039  & 0.000026 \\
196 & -9  & 11.996600 & 11.9965140 & 0.000086\\
20 & 13  & 4.353070 & 4.3530336 & 0.000036\\
130 & -22  & 10.051700 & 10.0516770 & 0.000023
\end{tabular}
\end{table}

One measure of the quality of the WKB approximation is a
comparison of the eigenenergies calculated with the RKF method and
the WKB approximation. Table \ref{energies} shows the eigenergies
for a few selected states. Since the WKB approximation should be
most accurate at large energies and angular momenta, it is not
surprising that the largest difference is for the ground state
(lowest positive energy state), although even here the difference
is less that 2 MeV. For the other states the difference is less
that 0.1 MeV.

\begin{figure}
\centerline{\includegraphics[height=4.0in]{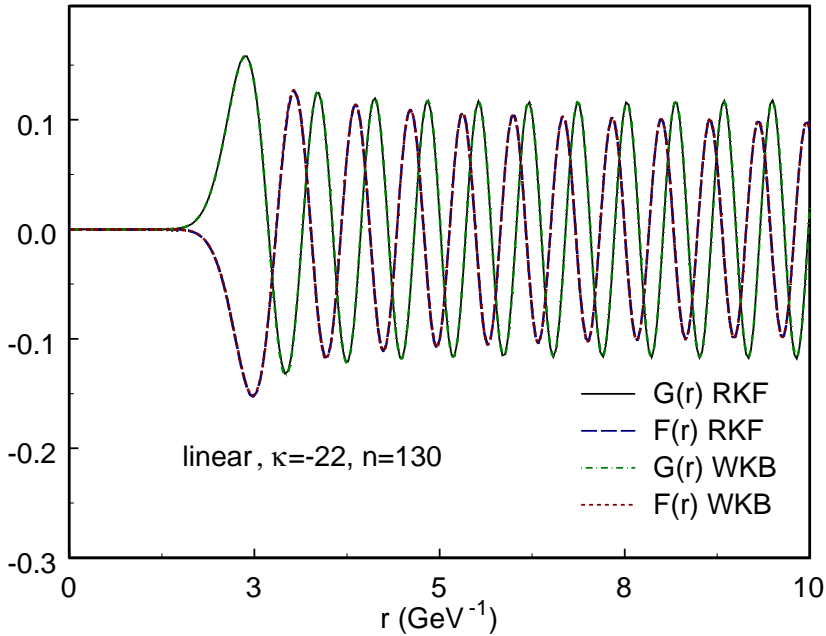}}
\caption{Reduced radial wave functions for $\kappa=-22$ and
$n=130$ for a linear confining potential. Upper and lower
components are shown for direct integration of the Dirac equation
using the Runge-Kutta-Fehlberg (RKF) method and in the WKB
approximation.}\label{excited}
\end{figure}

Figure \ref{excited} shows the upper and lower component wave
functions for $\beta=0$ with $\kappa=-22$, $n=130$ and $E=10.051\
{\rm GeV}$ (the last state listed in Table \ref{energies}). Wave
functions are shown for both direct integration of the Dirac
equation using the Runge-Kutta-Fehlberg (RKF) method and for the
WKB approximation. The range in $r$ is chosen to cover the region
of appreciable overlap between this state and the lowest positive
energy state. The differences between the two sets of wave
functions can not be distinguished in this figure. This indicates
that the Dirac WKB approximation is very good at these energies.

\begin{figure}
\centerline{\includegraphics[height=2.5in]{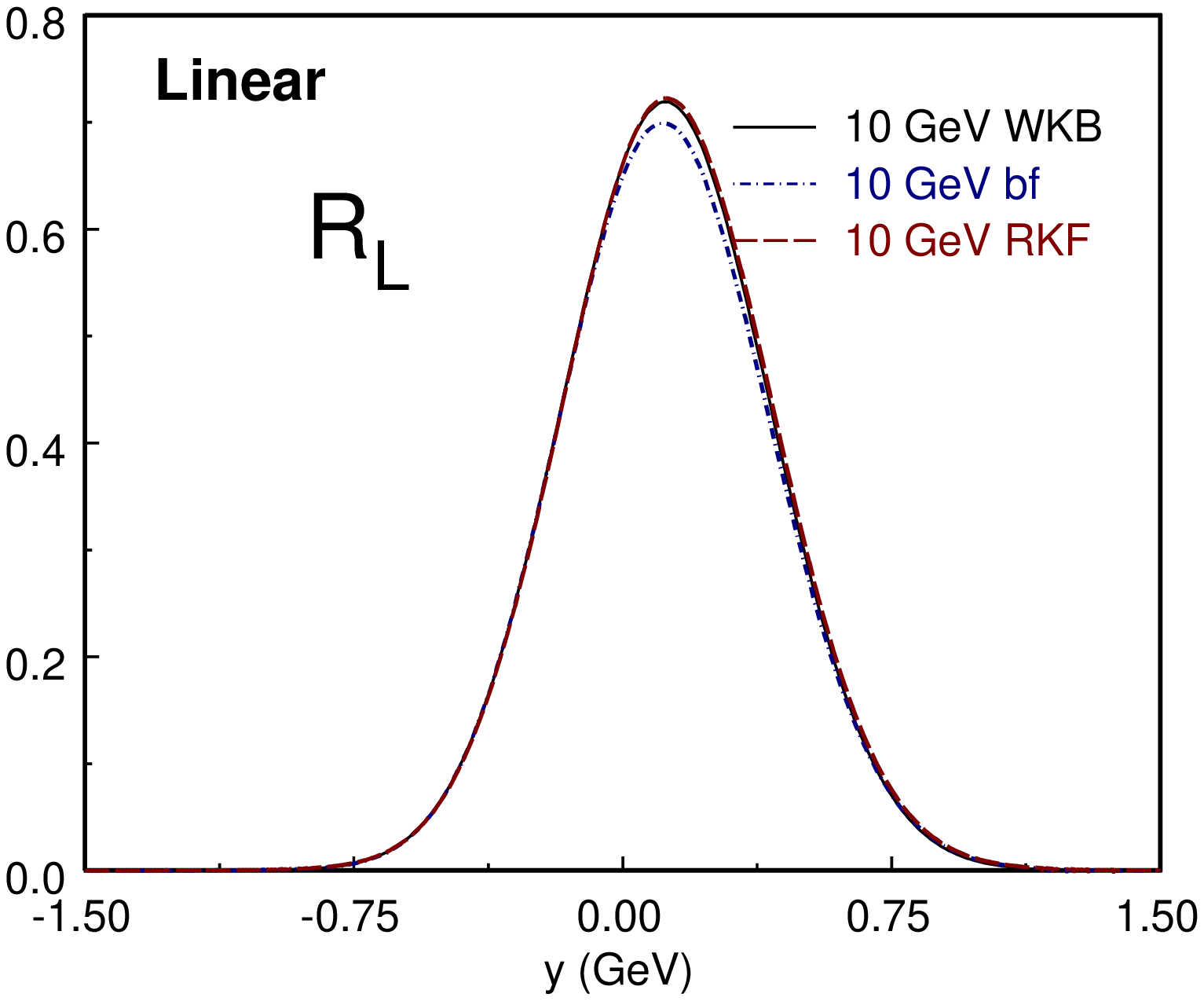}
\includegraphics[height=2.5in]{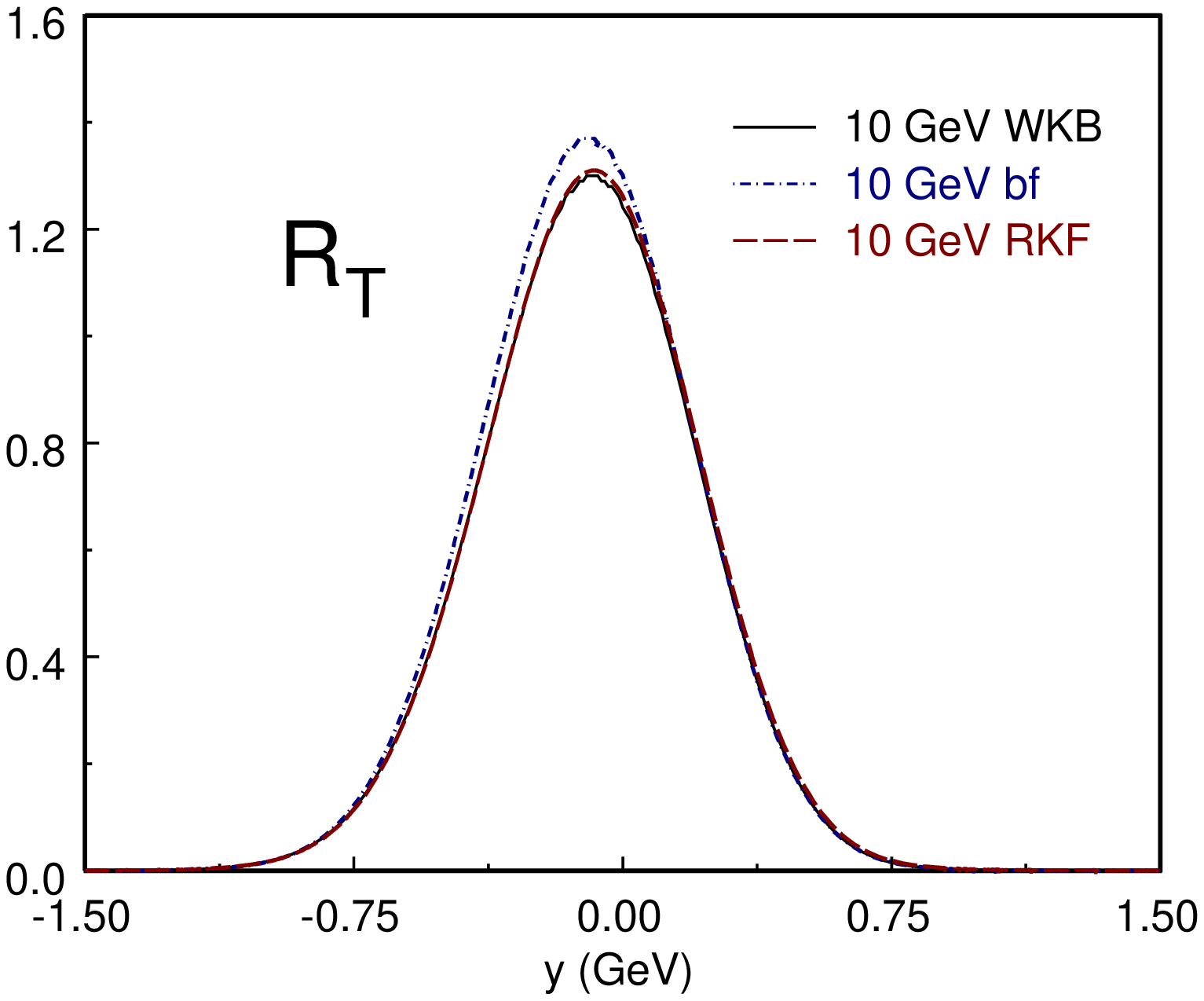}}
\centerline{\includegraphics[height=2.5in]{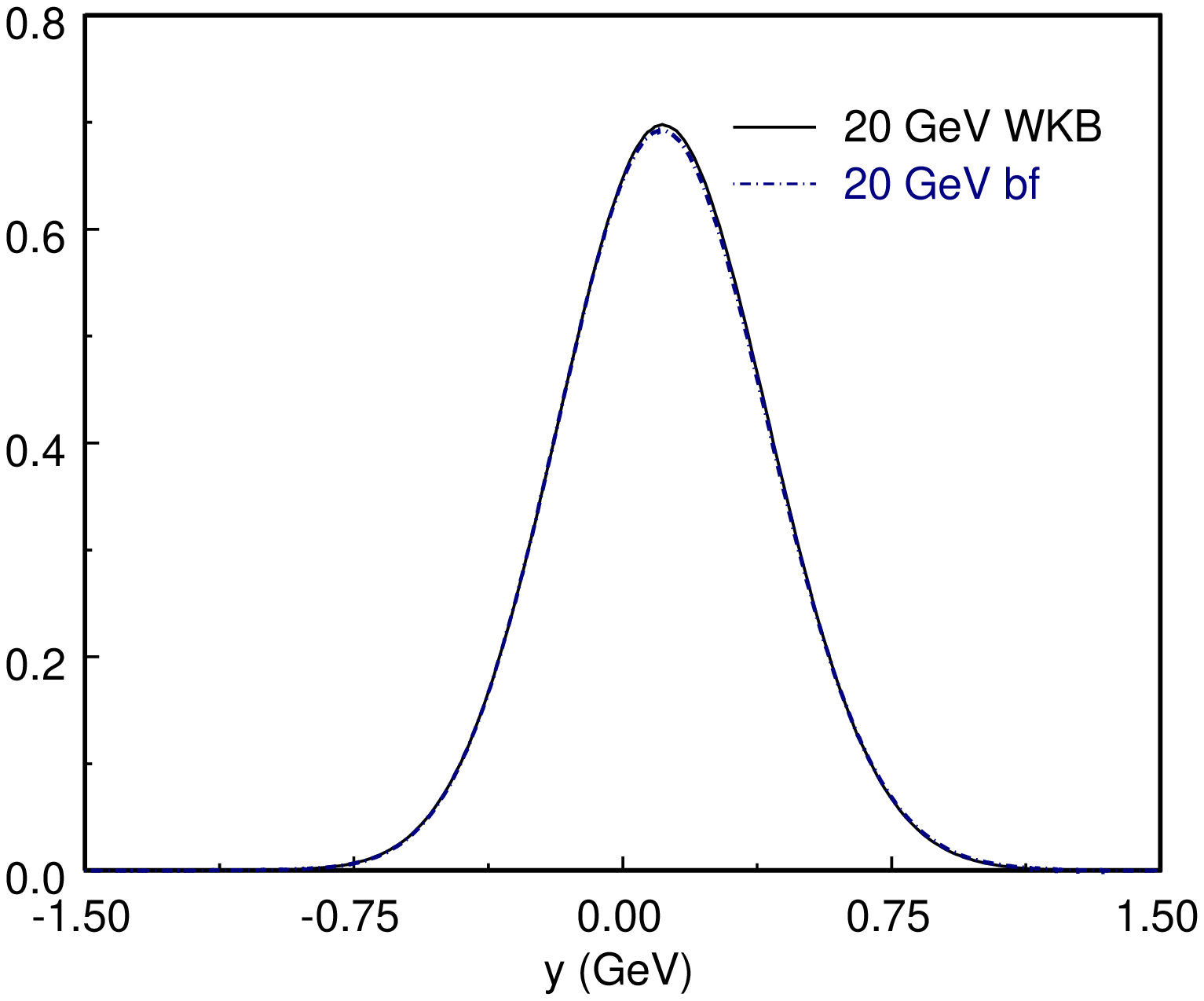}
\includegraphics[height=2.5in]{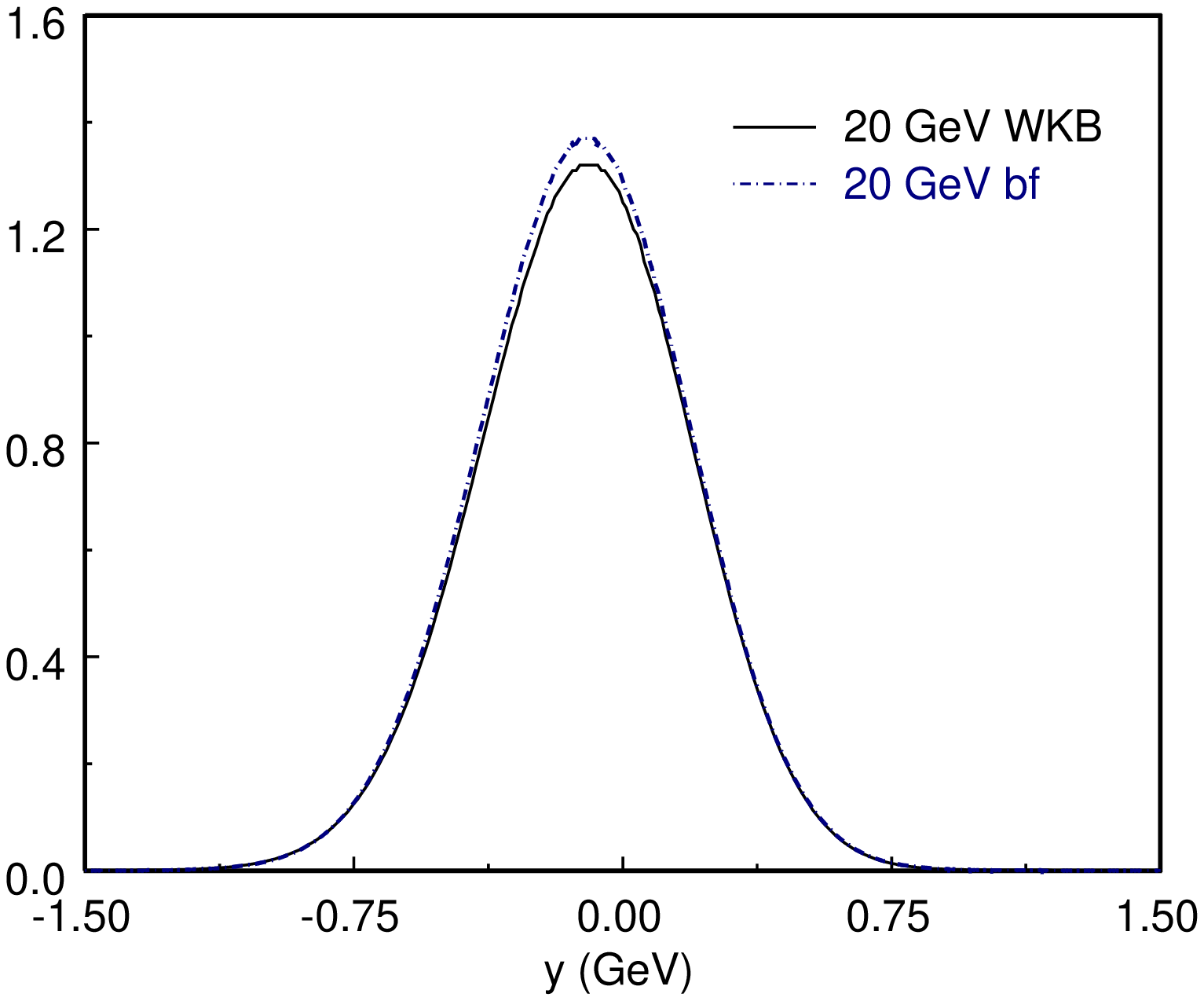}}
\centerline{\includegraphics[height=2.5in]{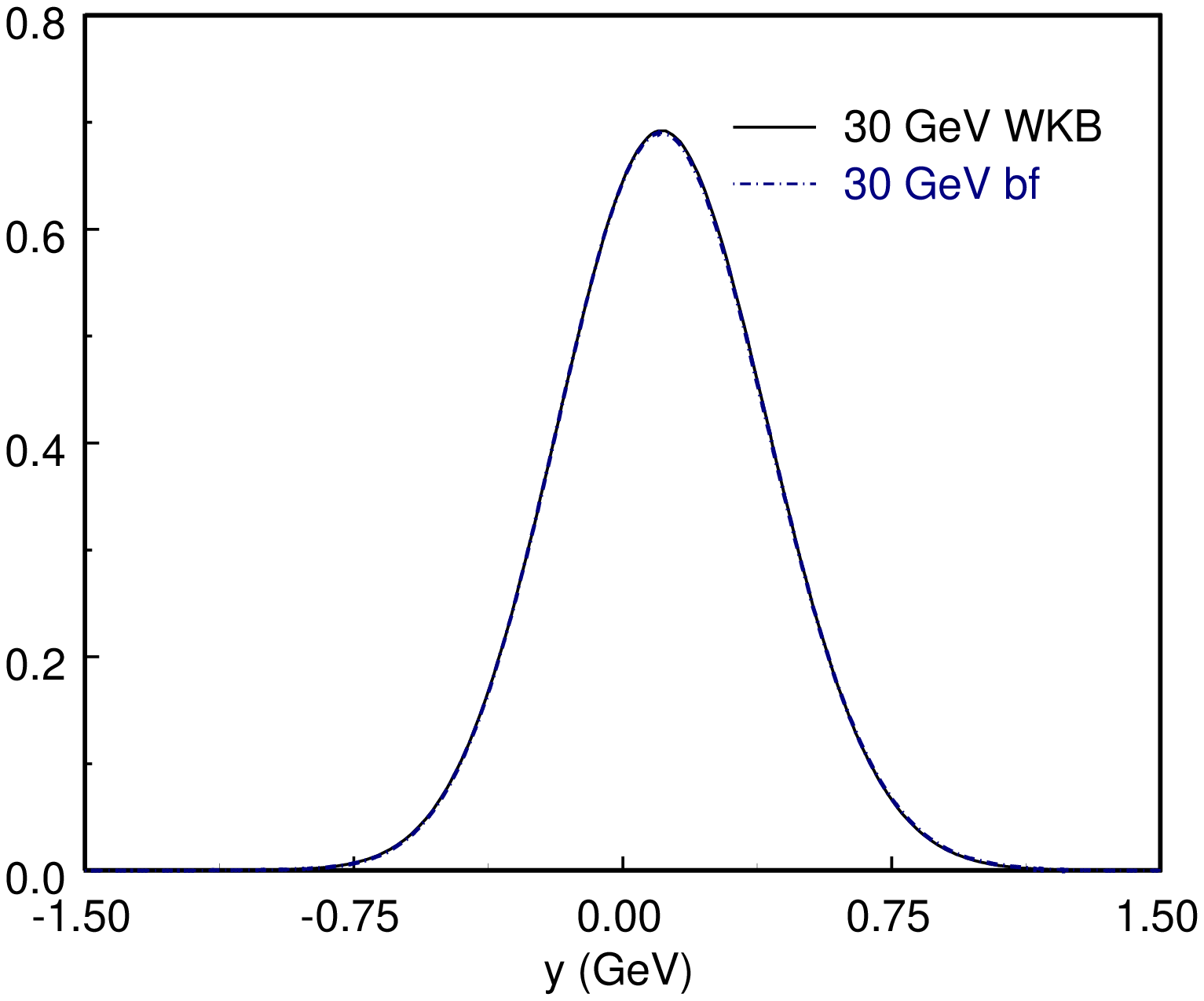}
\includegraphics[height=2.5in]{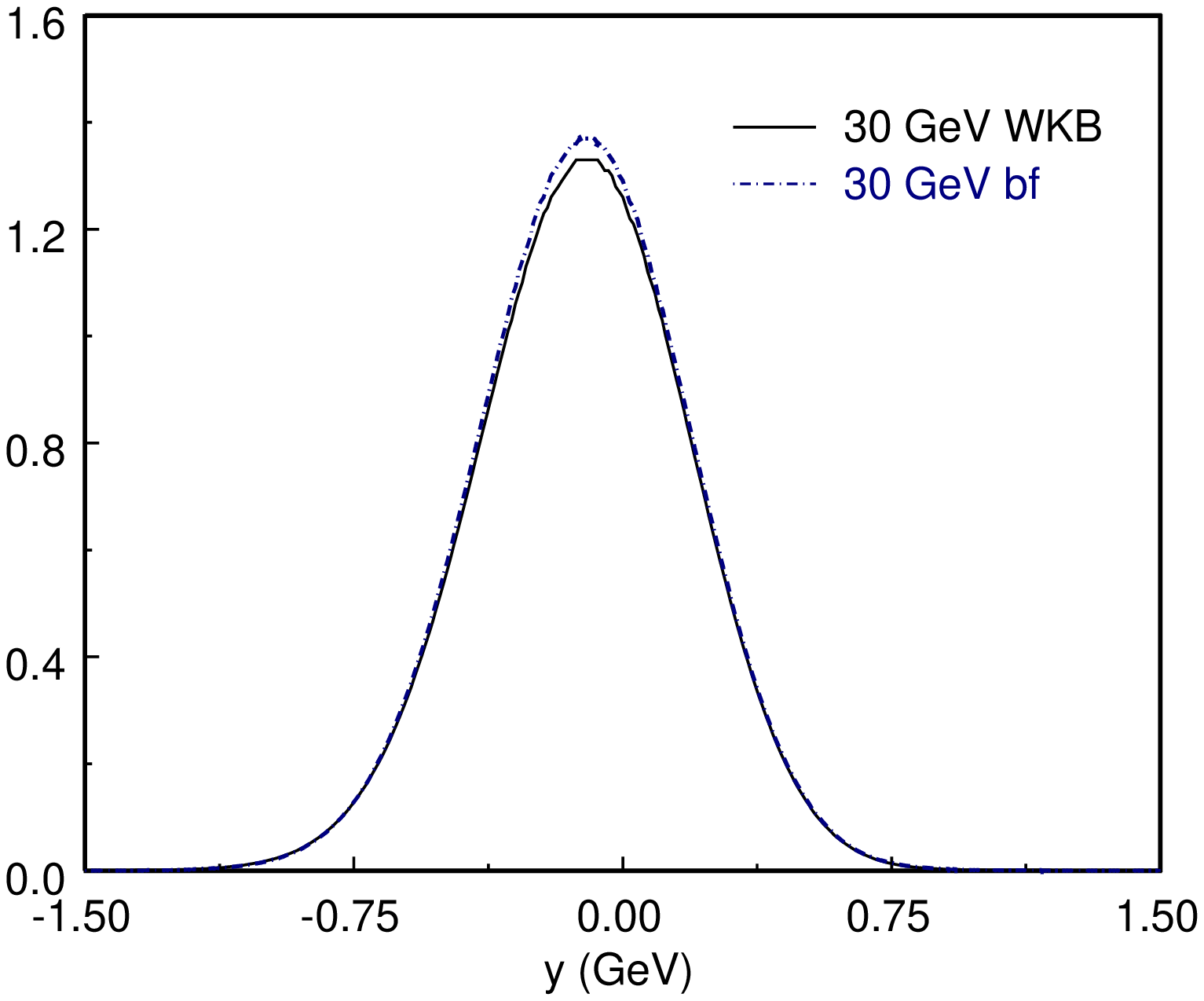}}
\caption{Longitudinal and transverse response functions for the
linear confining potential ($\beta=0$) at $q=$10, 20 and 30 GeV. At
each momentum transfer the bound-bound response function is compared
to the bound-free response. The corresponding responses using the
RKF integration are shown for comparison.}\label{linear1}
\end{figure}

Figure \ref{linear1} shows the longitudinal and transverse response
functions for three-momentum transfers of 10, 20 and 30 GeV as a
function of the y-scaling variable. Each plot contains a comparison
of the model with WKB wave functions to the response calculated
assuming that the quark is ejected from the bound state to a
plane-wave continuum state, the bound-free transition. This
bound-free result is the model equivalent of the parton model. In
addition the plots for $q=$ 10 GeV also show the model response
calculated with RKF wave functions. It is clear from these that the
RKF and WKB responses are virtually identical, as would be expected
from the comparison of wave functions in Fig. \ref{excited}. This
figure shows that model response functions approach the
corresponding bound-free response functions as the momentum transfer
increases and that at $q=$ 30 GeV they are almost identical. This
indicates that the bound-bound transition within a model with a
linear confining potential is dual to the bound-free case.

\begin{table}
\caption{Extrapolation of the peak value $R_{\rm max}$ and
position $y_{\rm max}$ of the peak of the response functions for
$q\rightarrow\infty$. $\Delta R$ is the difference in peak height
between the bound-free and bound-bound
calculations.}\label{asymptote}
\begin{tabular}{|c|r|r|r|r|r|r|}\hline\hline
& & \multicolumn{2}{c|}{bound-bound}&
\multicolumn{2}{c|}{bound-free}&
\\\hline
 & & $y_{\rm max}$ & \multicolumn{1}{c|}{$R_{\rm max}$} & $y_{\rm max}$ & $R_{\rm
max}$& $\Delta R$\\\tableline
linear & $R_L$ & 0.102 &  0.683   $\pm$   0.005  & 0.105 &  0.685 & 0.002\\
& $R_T$ & -0.106 & 1.358 $\pm$ 0.008 &  -0.105 & 1.370 &
0.012\\\hline
linear + Coulomb & $R_L$ & 0.150 &  0.606   $\pm$   0.003 & 0.132 & 0.614 & 0.008\\
& $R_T$ & -0.108 & 1.137   $\pm$   0.005 &  -0.132 & 1.229 &
0.092\\\hline
\end{tabular}
\end{table}
\begin{figure}
\centerline{\includegraphics[height=2.5in]{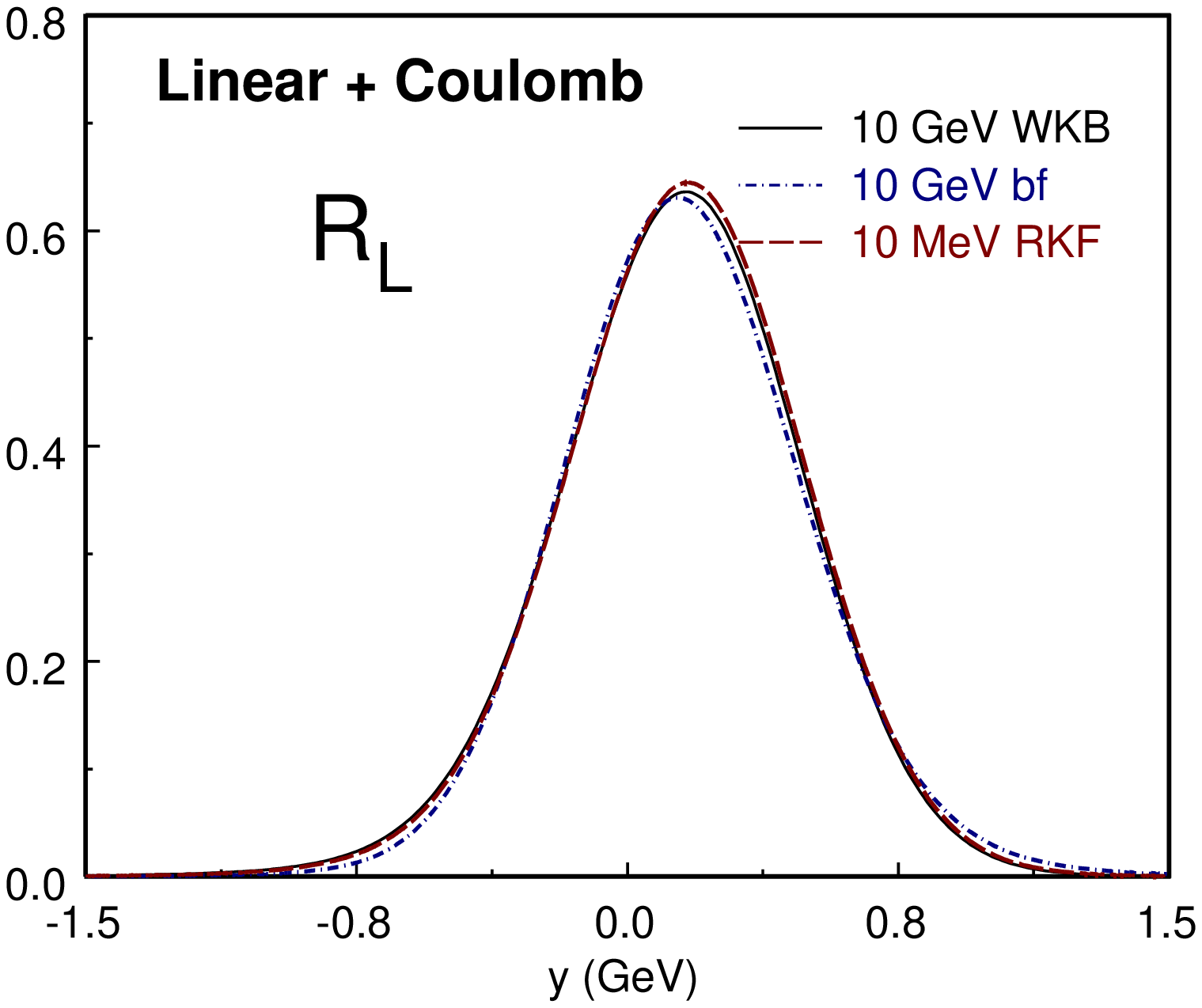}
\includegraphics[height=2.5in]{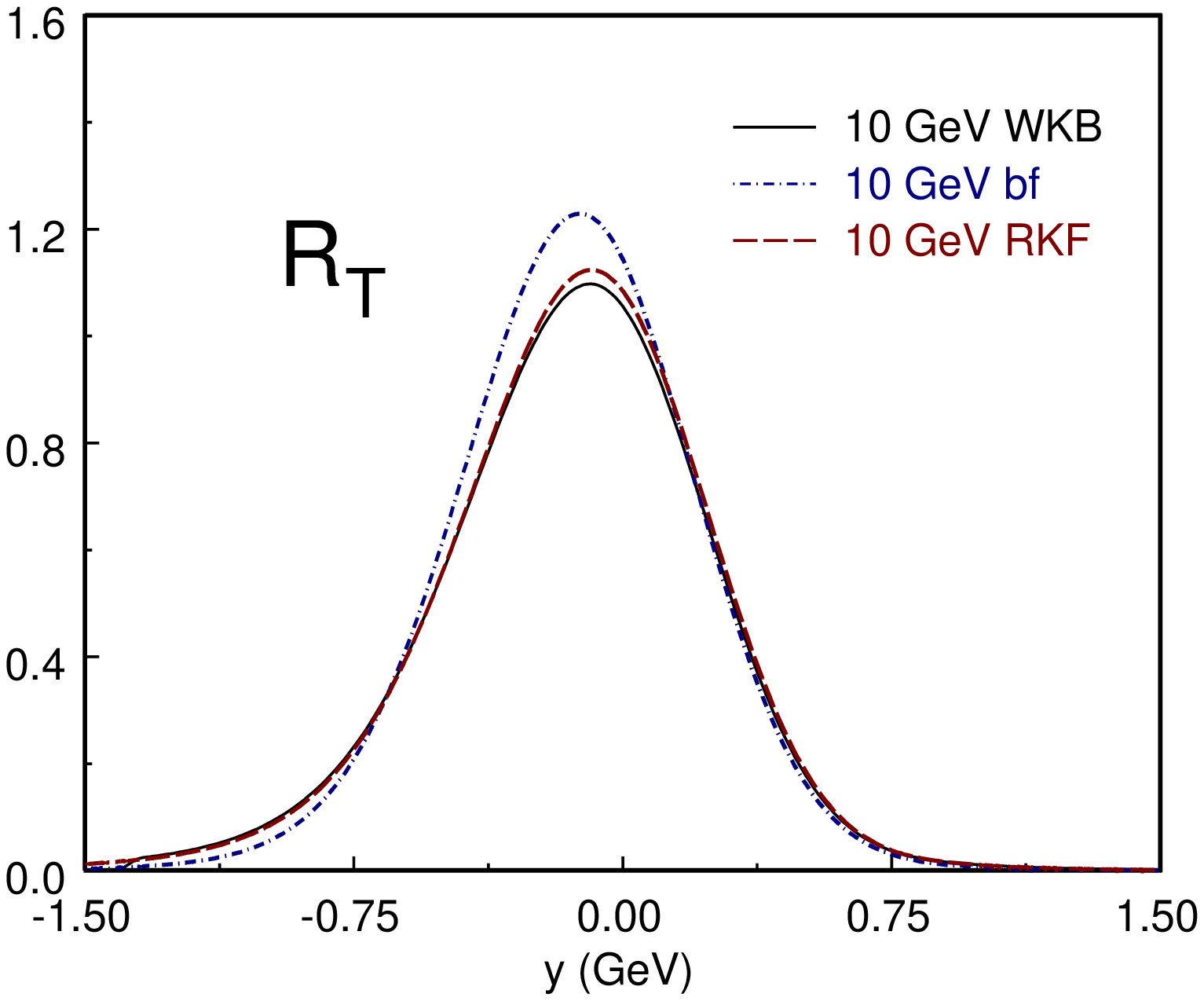}}
\centerline{\includegraphics[height=2.5in]{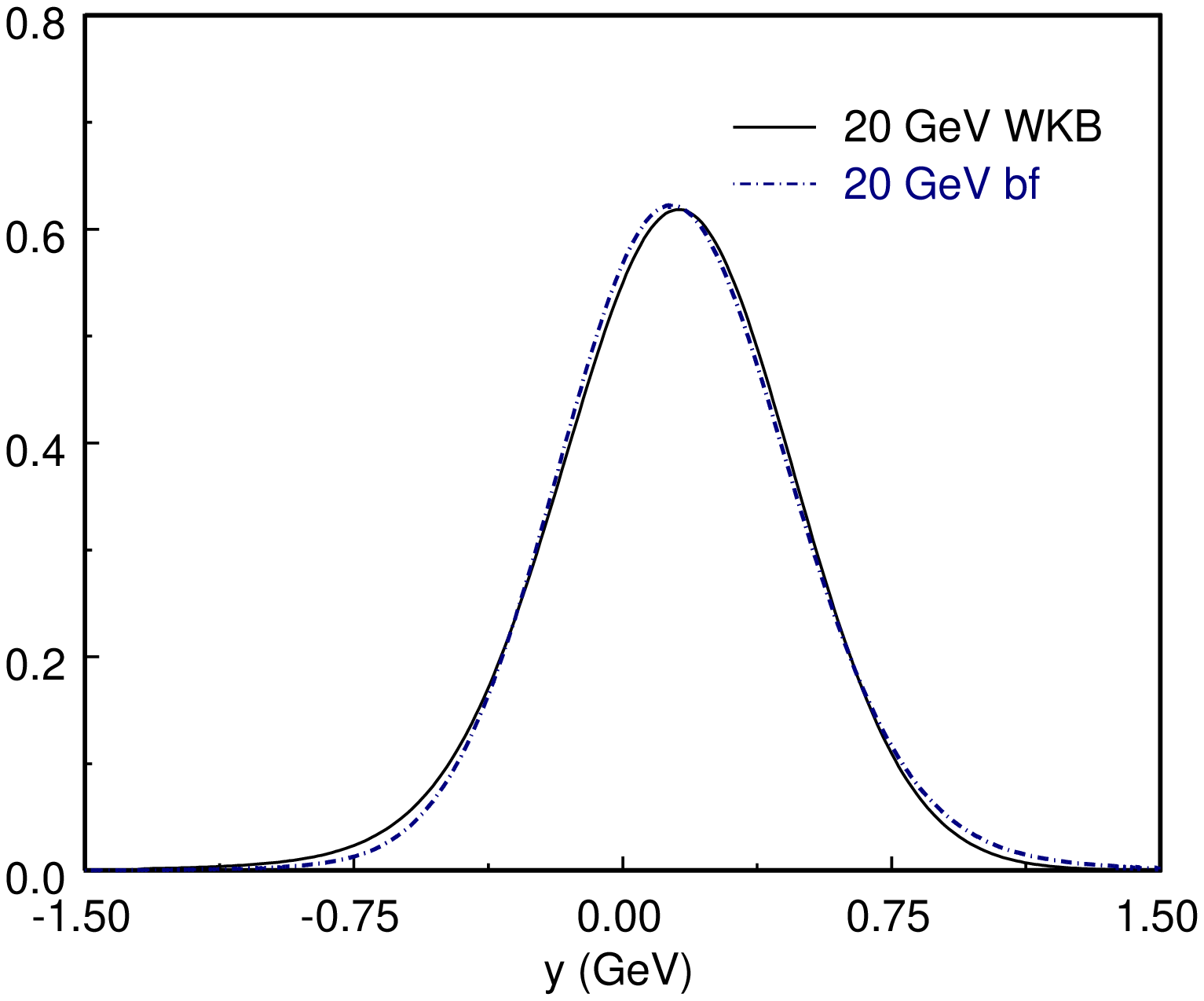}
\includegraphics[height=2.5in]{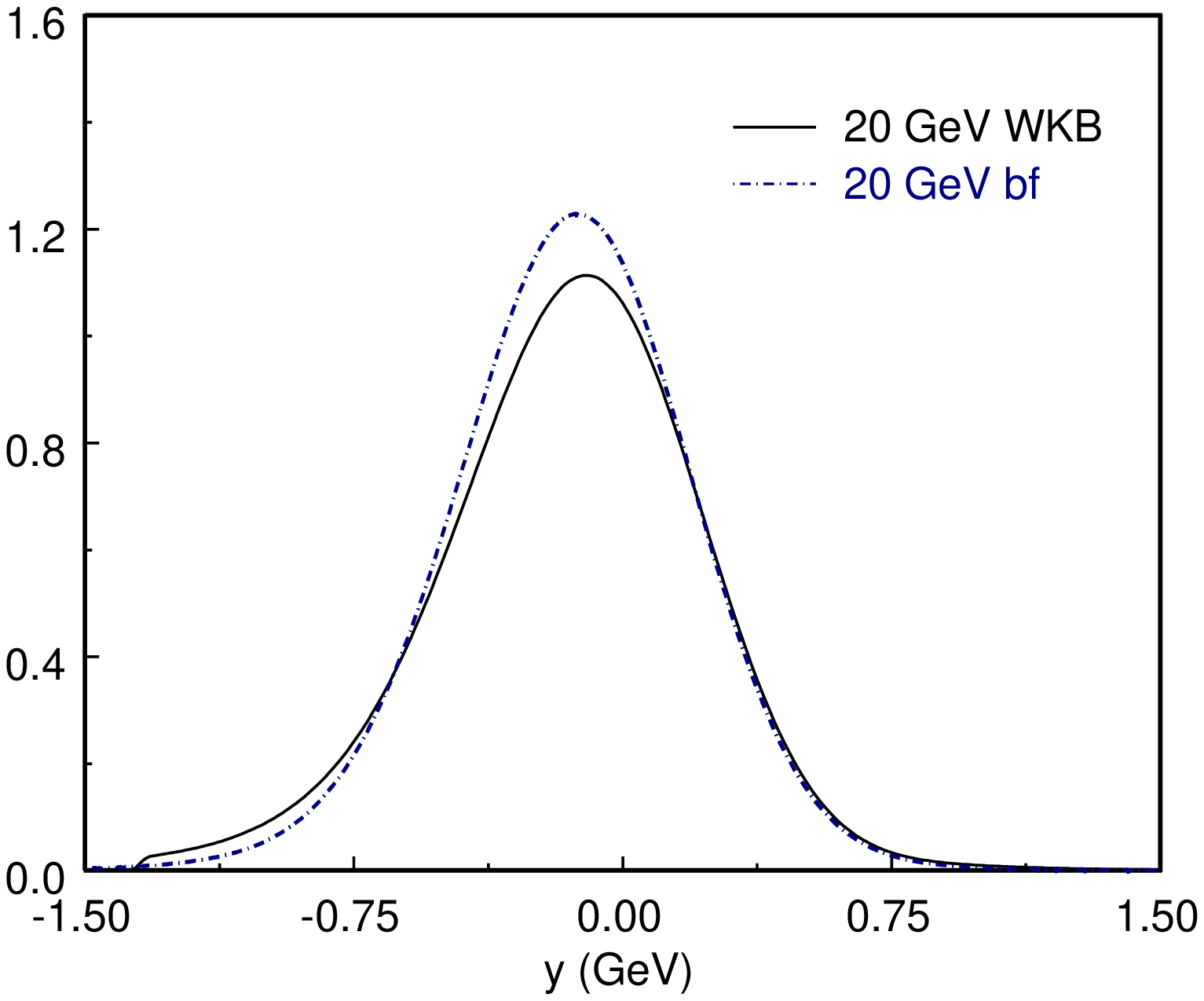}}
\centerline{\includegraphics[height=2.5in]{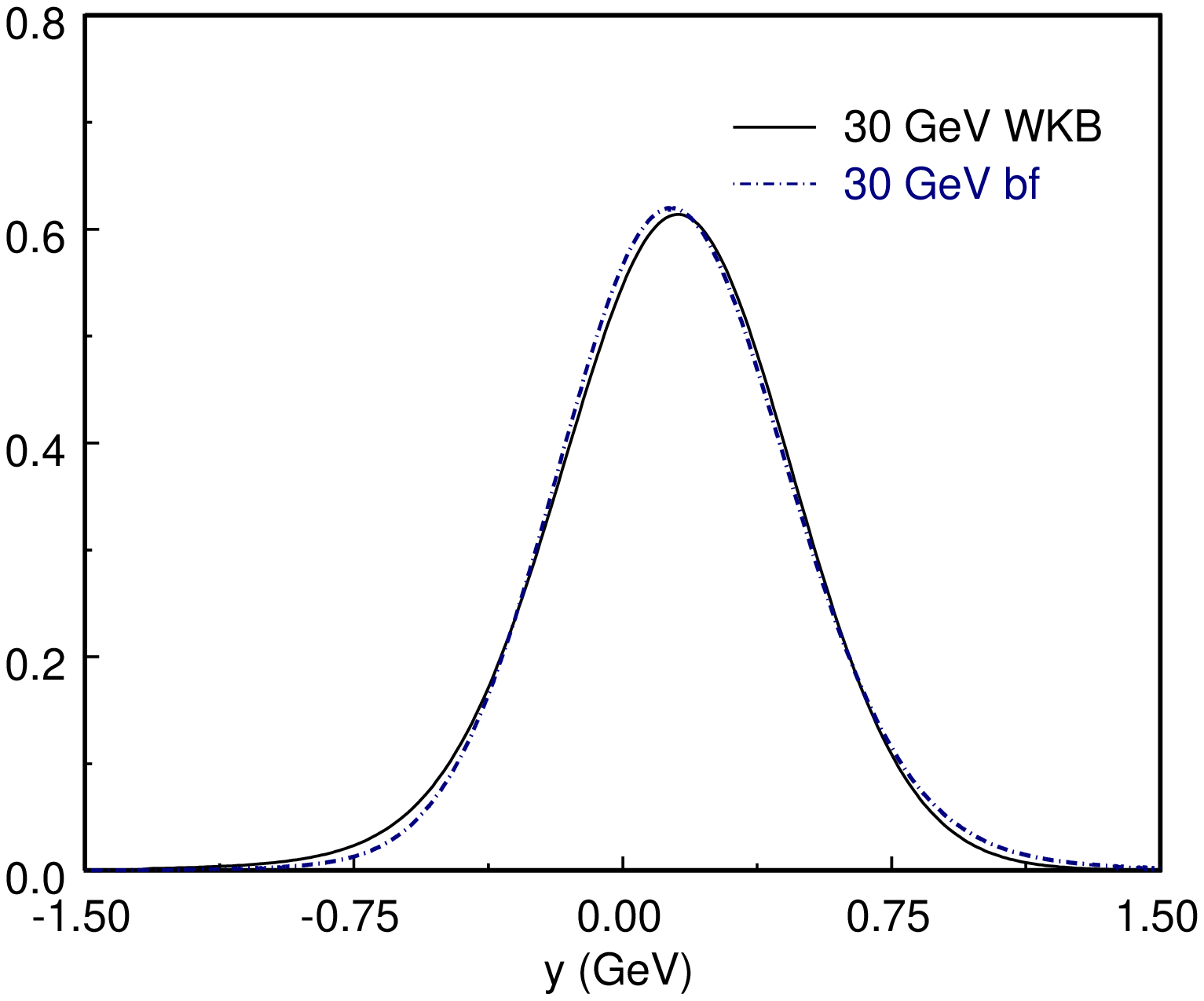}
\includegraphics[height=2.5in]{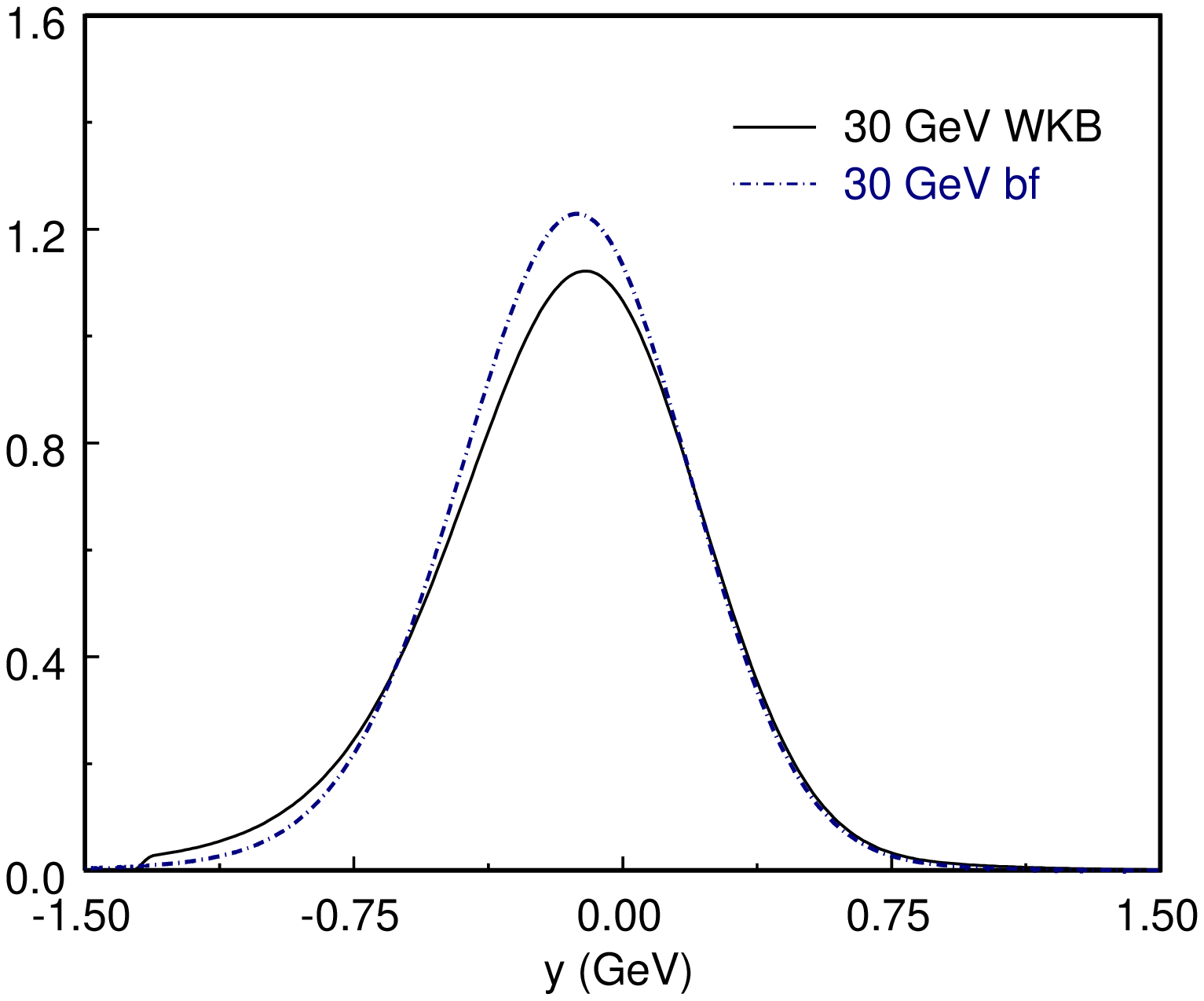}}
\caption{Longitudinal and transverse response functions for the
linear confining plus Coulomb potentials ($\beta=0.4$) at $q=$10,
20 and 30 GeV. At each momentum transfer the bound-bound response
function is compared to the bound-free (bf) response. The
corresponding responses using the RKF integration are shown for
comparison.}\label{LPC1}
\end{figure}

Figure \ref{LPC1} shows similar response functions for the case
where the vector Coulomb potential is included. Again, the RKF and
WKB response functions are very similar at $q=$ 10 GeV. As the
momentum transfer increases, the longitudinal model response for
the bound-bound transition appears to be approaching the
bound-free result from above.  For the transverse response, the
bound-bound response is substantially smaller than the bound-free.
Although the bound-bound response does slowly approach the
bound-free with increasing momentum transfer, it is still quite
far from the bound-free at $q=$ 30 GeV. Also note that for the
both longitudinal and transverse responses the shapes of the model
result and the bound-free results are slightly different and this
does not seem to be changing with momentum transfer. This seems to
indicate that the bound-bound and bound-free responses are not
dual once the Coulomb potential is introduced. An extrapolation of
the peak positions and values of the response functions has been
performed fitting the bound-bound calculations at 10, 15, 20, 25
and 30 GeV to the form
\begin{equation}
 R_{\rm max}(q)=A+\frac{B}{q}+\frac{C}{q^2}\,.
\end{equation}
An estimate of the extrapolation error was obtained by varying the
number of points in the fit from 3 to 5. Asymptotic values for the
bound-free case are obtained directly from a direct calculation of
the responses in the limit $q\rightarrow\infty$. The results of
the extrapolation are shown in Table \ref{asymptote}. For the case
of the linear potential the positions and peak values of the
bound-bound response functions do not very significantly from the
asymptotic bound-free response functions. However, for the
linear-plus-coulomb potential, there is a consistent shift in the
peak position and the difference for the height of the transverse
response is substantial. A similar situation has been reported
elsewhere for the case where the scalar and vector potentials are
assumed to be identical linear confining potentials
\cite{mpdirac,marka1}.

In this paper, we have presented a new method to solve the Dirac
equation for scalar and time-like vector potentials in the WKB
approximation. We have applied this method to calculating the
response functions of a light quark bound to an infinitely heavy
di-quark in the bound-bound transition at very high momentum
transfers. This type of calculation is relevant for modeling quark
hadron duality. We compared these results to the bound-free
transition, and found that the responses scale to the same limit for
just a scalar potential. The vector potential introduces small
differences in the two scaling functions.

\acknowledgments

Two of the authors (SJ and JT) thanks the Theory Group of
Jefferson Lab for their hospitality. This work was supported in
part by funds provided by the U.S. Department of Energy (DOE)
under cooperative research agreement under No. DE-AC05-84ER40150
and by the National Science Foundation under grants No.
PHY-0139973 and PHY-0354916.

\appendix

\section{The Dirac WKB wave functions near the turning points}

Since the WKB expansion does not converge near the classical turning
points, the WKB wave functions near these points are not a good
approximation to the exact wave functions and are, in fact, singular
at the turning points. This problem can be eliminated by matching
the WKB wave functions at some distance from the turning points to
an exact or approximate solution to the wave equations near the
turning points. The usual approach is to approximate the solution
near the turning points by assuming that the potential is roughly
linear over the region where the WKB wave functions fail. This gives
matching conditions that lead to the phase in (\ref{matchPhase}) and
the corresponding wave functions are Airy functions $Ai(z)$. The
matching to the WKB wave functions is achieved by using the limiting
properties of the Airy function
\begin{equation}
\lim_{z\rightarrow\infty}\pi^{\frac{1}{2}}
z^{\frac{1}{4}}Ai\left(z\right) \rightarrow\frac{1}{2}e^{-\zeta}
\label{Airy1}
\end{equation}
and
\begin{equation}
\lim_{z\rightarrow\infty}\pi^{\frac{1}{2}}
z^{\frac{1}{4}}Ai\left(-z\right)
\rightarrow\cos\left(\zeta-\frac{\pi}{4}\right)\,,\label{Airy2}
\end{equation}
where $\zeta=\frac{2}{3}z^{\frac{3}{2}}$ can be identified as the
WKB phase function. The factors of $z^{\frac{1}{4}}$ cancel the
singularity coming form the factor of $k_0^{-\frac{1}{2}}$ that
appears in the definition of the WKB wave functions and gives a
smooth finite result at the classical turning points.

A similar procedure can be used for the Dirac WKB wave functions
once a few additional complications are dealt with. The primary
problem is that the local wave vector $k_1(r)$ is singular at the
turning points. This is a result of the factor of $k_0(r)$ that
appears in (\ref{k1}). It is easy to show that near the turning
points $k_0(r)\sim |r-r_\pm|^\frac{1}{2}$. Therefore, $k_1(r)\sim
|r-r_\pm|^{-\frac{1}{2}}$. Since this singularity is integrable,
$\Re \xi_1(r)$ will be finite at the turning points but will have
infinite slope. This phase also changes sign at these points. One
result of this is to cause the total phase of the WKB wave
functions to have an additional zero near the classical turning
point. This interferes with the use of the usual Airy function
approximation to the wave functions near the turning point for the
upper component wave function.

For the lower component wave function, there is the additional
complication that it contains an explicit factor of $k_1(r)$ and is
therefore more singular at the turning point than can be cancelled
by the Airy function approximation, provided that a function that
smoothly extrapolates to the sine function can be constructed. There
is an additional problem arising from the factor of $k_0'(r)/k_0(r)$
that appears in (\ref{Imxi1}). This behaves like
$|r-r_\pm|^{-\frac{3}{2}}$ near the turning points and cannot be
cancelled by the Airy function solutions.

Consider the solution in Region II given by (\ref{GII}). Define
the phases
\begin{eqnarray}
\zeta^I_i(r)&=&\int_r^{r_-} dr'
\tilde{k}_i(r')\\
\zeta^{II-}_i(r)&=&\int_{r_-}^r dr'
k_i(r')\\
\zeta^{II+}_i(r)&=&\int_{r}^{r_+} dr'
k_i(r')\\
\zeta^{III}_i(r)&=&\int_{r_+}^{r} dr' \tilde{k}_i(r')
\end{eqnarray}
where $i=0,1$. Equation (\ref{GII}) can then be written as
\begin{eqnarray}
G_{II}(r)&=&N\frac{\sqrt{m+S(r)-V(r)+E}}{\sqrt{k_0(r)}}\cos\left(\zeta^{II-}_0(r)+\zeta^{II-}_1(r)
-\frac{\pi}{4}\right)\nonumber\\
&=&N\frac{\sqrt{m+S(r)-V(r)+E}}{\sqrt{k_0(r)}}\left(\cos\left(\zeta^{II-}_0(r)
-\frac{\pi}{4}\right)\cos(\zeta^{II-}_1(r))\right.\nonumber\\
&&\qquad\left.-\sin\left(\zeta^{II-}_0(r)
-\frac{\pi}{4}\right)\sin(\zeta^{II-}_1(r))\right)
\end{eqnarray}
Near the lower classical turning point, $\cos\left(\zeta^I_0(r)
-\frac{\pi}{4}\right)$ can be replaced by the Airy function using
(\ref{Airy2}) and $\sin\left(\zeta^I_0(r) -\frac{\pi}{4}\right)$
can be replaced using the identity
\begin{equation}
\lim_{z\rightarrow\infty}\pi^{\frac{1}{2}}
z^{-\frac{1}{4}}Ai'\left(-z\right)
\rightarrow\sin\left(\zeta-\frac{\pi}{4}\right)\,.\label{Airy3}
\end{equation}
Therefore, near the lower classical turning point in Region II we
replace the WKB wave function (\ref{GII}) by
\begin{eqnarray}
G_{II}(r)&\rightarrow&N\frac{\sqrt{m+S(r)-V(r)+E}}{\sqrt{k_0(r)}}\pi^{\frac{1}{2}}
z^{\frac{1}{4}}(\zeta^{II-}_0(r))\nonumber\\
&&\times\left(Ai\left(-z(\zeta^{II-}_0(r))
\right)\cos(\zeta^{II-}_1(r))
-\frac{1}{\sqrt{z(\zeta^{II-}_0(r))}}Ai'\left(-z(\zeta^{II-}_0(r))
\right)\sin(\zeta^{II-}_1(r))\right)\nonumber\\  \, . \label{GIIa}
\end{eqnarray}
This can be continued across the lower turning point into Region I
using (\ref{Airy1}) and the identity
\begin{equation}
-\lim_{z\rightarrow\infty}\pi^{\frac{1}{2}}
z^{-\frac{1}{4}}Ai'\left(z\right) \rightarrow\frac{1}{2}e^{-\zeta}
\label{Airy4}
\end{equation}
to give
\begin{eqnarray}
G_{I}(r)&\rightarrow&N\frac{\sqrt{m+S(r)-V(r)+E}}{\sqrt{k_0(r)}}\pi^{\frac{1}{2}}
z^{\frac{1}{4}}(\zeta^{I}_0(r))\nonumber\\
&&\times\left(Ai\left(z(\zeta^{I}_0(r))
\right)\cosh(\zeta^{I}_1(r))
+\frac{1}{\sqrt{z(\zeta^{I}_0(r))}}Ai'\left(z(\zeta^{I}_0(r))
\right)\sinh(\zeta^{I}_1(r))\right)\nonumber\\\label{GIa}
\end{eqnarray}
These wave functions are now continuous and smooth at the lower
classical turning point, and match (\ref{GII}) and (\ref{GI}) away
from the turning points. The solutions near the upper turning
point are obtained by the replacement
$\zeta^{II-}_i(r)\rightarrow\zeta^{II+}_i(r)$ in (\ref{GIIa}) and
for Region III by the replacement
$\zeta^{I}_i(r)\rightarrow\zeta^{III}_i(r)$ in (\ref{GIa}).

The corresponding expressions for the lower components can be
obtained by using the expressions above in
\begin{equation}
F(r)=\frac{G'(r)+\frac{\kappa}{r}G(r)}{(m+S(r)-V(r)+E)}\,.
\end{equation}

\begin{figure}
\centerline{\includegraphics[height=4in]{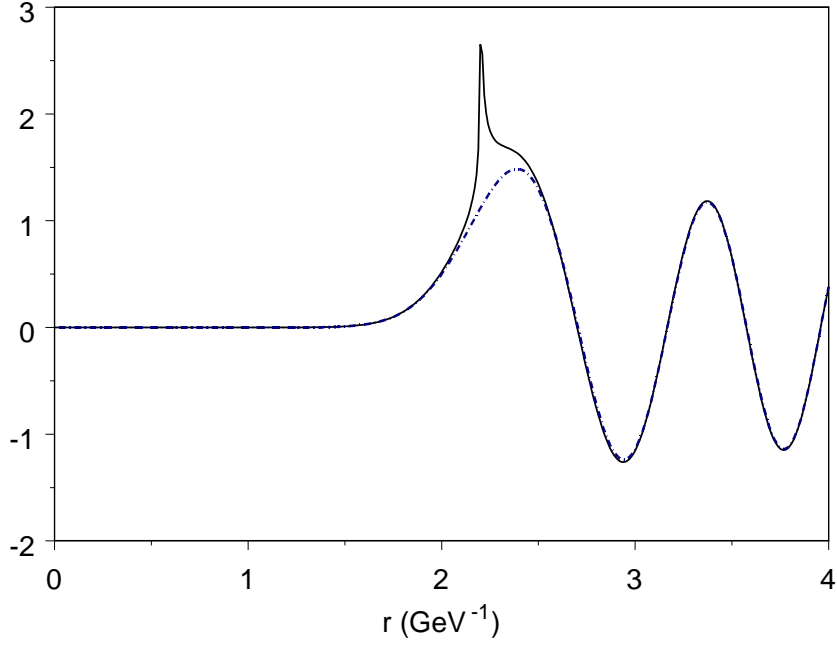}}
\caption{The WKB wave function is shown as solid line. The Airy
function, which smoothly interpolates between the WKB wave
functions in regions I and II, is shown as dash-dotted line.}
\label{figwfmatch}
\end{figure}

An example for the procedure outlined above is shown in Fig.
\ref{figwfmatch}. The solid line shows the WKB wave function given
by (\ref{GI}) and (\ref{GII}), which clearly becomes singular at the
turning point. The dash-dotted line is calculated using (\ref{GIa})
and (\ref{GIIa}). The interpolation of the wave function using the
Airy function clearly matches the WKB wave function within one wave
length on either side of the classical turning point. Equations
(\ref{GI}) and (\ref{GII}) can then be used to interpolate the WKB
wave function through the region of the classical turning point
where the WKB approximation is invalid.

\end{document}